\documentclass[
  journal=largetwo,
  manuscript=article-type,
  year=2020,
  volume=37
]{cup-journal}

\usepackage{aas-macros}
\usepackage[utf8]{inputenc}
\usepackage{csquotes}
\usepackage[english]{babel}

\usepackage{amsmath,amssymb}
\usepackage[nopatch]{microtype}
\usepackage{booktabs}
\usepackage{siunitx}
\usepackage{graphicx}
\usepackage{rotating}
\usepackage{multirow}
\usepackage{tablefootnote}
\usepackage{tikz}
\usepackage{upgreek}

\usepackage{biblatex}
\newcommand{\DMobs}{\ensuremath{{\rm DM}_{\rm obs}}}

\newcommand{\zdm}{{\sc zDM}}

\newcommand{\wi}{\ensuremath{w_i}} 
\newcommand{\wtau}{\ensuremath{w_\tau}} 
\newcommand{\wdm}{\ensuremath{w_{\rm DM}}} 
\newcommand{\tres}{\ensuremath{t_{\rm res}}} 
\newcommand{\weff}{\ensuremath{w_{\rm eff}}} 
\newcommand{\wapp}{\ensuremath{w_{\rm app}}} 
\newcommand{\wsnr}{\ensuremath{w_{\rm snr}}} 
\newcommand{\wobs}{\ensuremath{w_{\rm obs}}}
\newcommand{\wmax}{\ensuremath{w_i^{\rm max}}}
\newcommand{\taumax}{\ensuremath{\tau_{\rm obs}^{\rm max}}}
\newcommand{\hwmax}{\ensuremath{w_{i,\rm host}^{\rm max}}}
\newcommand{\whost}{\ensuremath{w_{i,\rm host}}}
\newcommand{\htaumax}{\ensuremath{\tau_{\rm host,1\, GHz}^{\rm max}}}

\newcommand{\taughz}{\ensuremath{\tau_{\rm 1\,GHz}}}
\newcommand{\tauhost}{\ensuremath{\tau_{\rm host,1\,GHz}}}
\newcommand{\tauobs}{\ensuremath{\tau_{\rm obs}}}

\newcommand{\pccc}{\si{pc\,\centi\metre^{-3}}}

\newcommand{\nsfr}{\ensuremath{n_{\rm sfr}}} 

\title{Estimation of intrinsic fast radio burst width and scattering distributions from CRAFT data}

\author{C.~W.~James}
\affiliation{International Centre for Radio Astronomy Research, Curtin University, Bentley, 6102, WA, Australia}
\email[Clancy W.\ James]{clancy.james@curtin.edu.au}

\author{J.~Hoffmann}
\affiliation{International Centre for Radio Astronomy Research, Curtin University, Bentley, 6102, WA, Australia}
\email[Clancy W.\ James]{clancy.james@curtin.edu.au}

\author{J.~X.~Prochaska}
\affiliation{Department of Astronomy and Astrophysics, University of California, Santa Cruz, CA 95064, USA}
\alsoaffiliation{Kavli Institute for the Physics and Mathematics of the Universe, 5-1-5 Kashiwanoha, Kashiwa 277-8583, Japan}
\alsoaffiliation{Division of Science, National Astronomical Observatory of Japan, 2-21-1 Osawa, Mitaka, Tokyo 181-8588, Japan}

\author{M.~Glowacki}
\affiliation{Institute for Astronomy, University of Edinburgh, Royal Observatory, Edinburgh, EH9 3HJ, United Kingdom}
\alsoaffiliation{International Centre for Radio Astronomy Research, Curtin University, Bentley, 6102, WA, Australia}
\alsoaffiliation{Inter-University Institute for Data Intensive Astronomy, Department of Astronomy, University of Cape Town, Cape Town, South Africa}

\keywords{radio transient sources, radio bursts} 

\begin{document}

\begin{abstract}
The intrinsic width and scattering distributions of fast radio bursts (FRBs) inform on their emission mechanism and local environment, and act as a source of detection bias and, hence, an obfuscating factor when performing FRB population and cosmological studies. Here, we utilise a sample of 29 FRBs with measured high-time-resolution properties and known redshift, which were detected using the Australian Square Kilometre Array Pathfinder (ASKAP) by the Commensal Real-time ASKAP Fast Transients Survey (CRAFT), to model these distributions. Using this sample, we estimate the completeness bias of intrinsic width and scattering measurements, and fit the underlying, de-biased distributions in the host rest-frame. In no case do our model fits prefer a down-turn at high values of the intrinsic distributions of either parameter in the 0.01--40\,ms range probed by the data.
Rather, when assuming a spectral scattering index of $\alpha = -4$, we find that the intrinsic scattering distribution at 1\,GHz is consistent with a log-uniform distribution above 0.04\,ms, and that this functional form is strongly favoured over the lognormal descriptions used by previous works.  We also find that the intrinsic width distribution rises as a Gaussian in log-space in the $0.03-0.3$\,ms range, with a log-uniform distribution above that slightly preferred to a lognormal distribution.
This confirms previous works suggesting that FRB observations are currently strongly width- and scattering-limited, and we encourage FRB searches to be extended to higher values of time-width. It also implies a bias in FRB host galaxy studies, although the form of that bias is uncertain.  Finally, we find that our updated width and scattering models --- when implemented in the {\sc zDM} code --- produces $\sim$10\% more FRBs at redshift $z=1$ than at $z=0$ when compared to alternative width/scattering models, highlighting that these factors are important to understand when performing FRB population modelling.
\end{abstract}

\section{Introduction}
\label{sec:intro}

Fast radio bursts (FRBs) are extragalactic transients that allow the study of ionised media through which they propagate \citep{lorimer_bright_2007, thornton_population_2013, Macquart2020}. While dispersion measure (DM) is the most common property analysed, analysis of FRB time-profiles also yields the total FRB width, which is commonly decomposed into an intrinsic width in the observer frame, $w_{\rm obs}$, and a width due to scattering in turbulent plasmas along the line-of-sight, $w_\tau$ \citep{Cordes_McLaughlin_2003}, which typically manifests as an exponential scattering tail. The nature of scattering is particularly important for both FRB progenitor studies and population modelling. Scattering has been used to identify turbulent gas in the vicinity of FRB progenitors \citep[e.g.][]{2015Natur.528..523M,sammons_two-screen_2023,2025A&A...693A.279P}, and has been suggested to be a biasing factor in the FRB host galaxy distribution \citep{bhandari_host_2020}. In FRB population modelling, scattering is a nuisance parameter since, along with the intrinsic FRB width distribution and DM smearing, it affects instrumental biases \citep{Connor2019}.  Disentangling the intrinsic FRB scattering distribution from observations is therefore relevant to a wide range of FRB studies.

Fitting of the intrinsic FRB scattering and width distributions began with a model of the apparent FRB width, \wapp\ --- i.e., without separately resolving the width into scattering and intrinsic width --- used in \citet{2021MNRAS.501.5319A}. Based on FRB observations at Murriyang (Parkes) and ASKAP within the 1000--1500\,MHz frequency range at $\sim$ms time-resolution, it described a total width distribution as a lognormal. The mean $\mu_{\rm app}$ and standard deviation $\sigma_{\rm app}$ of $\log_{10} w_{\rm app} [{\rm ms}]$ were found to be 0.427 and 0.90, respectively (throughout this work, we describe distributions of the logarithm base 10 of width and scattering in units of ms).
This model was implemented in the {\sc zDM} code, where it was shown that it implied an intrinsic apparent width distribution of $\mu_{\rm app},\sigma_{\rm app} = 0.74, 1.07$
once bias effects were accounted-for \citep{james_zdm_2022}. The model included no redshift dependence, since it was unclear whether the $(1+z)^{-3}$ suppression of scattering, or $1+z$ dilation of instrinsic width, would be dominant at high redshifts. Since then, the most comprehensive model of intrinsic FRB behaviour was developed by \citet{chimefrb_collaboration_first_2021}, who used a pulse-injection system to account for experimental bias effects \citep{2023AJ....165..152M}. Using lognormal distributions to model intrinsic FRB width and scattering distributions at 600\,MHz, they found $\mu_w,\sigma_w = 0.0,0.42$, and $\mu_\tau,\sigma_\tau = 0.30,0.75$. The authors however note that there is little evidence for a down-turn in the scattering distribution above $10$\,ms, urge caution in interpreting results in the width distribution due to limitations with pulse injection, and present jacknife studies suggesting unmodelled correlations between parameters. Furthermore, since the majority of the CHIME sample had unknown redshifts, these values could not be corrected to the host rest-frame.

We revisit this question for three reasons. Firstly, new high-time-resolution data has become available from both the Canadian Hydrogen Intensity Mapping Experiment \citep[CHIME;][]{CHIME_baseband_2024} and the Australian Square Kilometre Array Pathfinder \citep[ASKAP; ][]{hotan_australian_2021,CRAFT_HTR}, revealing that the observed scattering times seen by CHIME at 600\,MHz \citep{CHIME_baseband_morphology_2025} are essentially identical to those observed by ASKAP at 1\,GHz \citep{CRAFT_HTR}. Those authors suggest that, given the differences in observation parameters between these two instruments, the observed scattering distributions would also have been expected to differ, unless, perhaps, both samples are completely dominated by aforementioned experimental bias effects. Secondly, the observed correlation between scattering and DM for Galactic pulsars \citep[e.g.][]{Bhat2004} should imply a similar correlation for FRB host galaxies, allowing scattering to constrain the host DM and, hence, improve redshift estimation for unlocalised FRBs \citep{CordesTauRedshift2022}. However, no correlations between either total DM \citep{2022ApJ...927...35C}, or excess DM \citep{CRAFT_HTR}, are evident, despite evidence that significant FRB scattering is indeed dominated by the host galaxy \citep{2022MNRAS.514.5866G,sammons_two-screen_2023}. \citet{2026ApJ...998....1M} suggest that fluctuations in the DM due to intervening intergalactic and circumgalactic media, combined with experimental bias effects in measuring $\tau$, may be the reason behind such a lack of correlation --- further emphasising the need to treat experimental biases.
Our third reason is concern between the potential influence of assumed scattering/width behaviour and estimates of cosmological source evolution.

This paper is therefore laid out as follows. In \S \ref{sec:method}, we outline the methods by which we analyse the intrinsic scattering distribution of FRBs. In \S \ref{sec:fits}, we present updated estimates of the intrinsic FRB scattering and width distributions, accounting for selection effects, and test lognormal distributions against other functional forms. In \ref{sec:zdm}, we use the \zdm\ code to estimate the effect of different width/scattering models on simulations of the FRB population.
We further discuss implications of our work in \S\ref{sec:discussion}. Throughout, we concentrate on the behaviour of these distributions at high time widths, which is both where experimental bias effects become important, and where fits to high-time-resolution FRB data become more reliable.

\section{Method}
\label{sec:method}

\subsection{Data}
\label{sec:data}

For this study, we use the high-time-resolution sample of FRBs published by \citet{CRAFT_HTR}, which were detected by ASKAP in incoherent sum \citep[ICS; ][]{Shannon_ICS} mode by the Commensal Real-time ASKAP Fast Transients \citep[CRAFT; ][]{MacquartCRAFT} Collaboration. Offline processing ---- described by \citet{scott_celebi_2023} --- coherently dedispersed these FRBs and formed tied beams at the FRB location, producing high signal-to-noise samples from which both intrinsic widths and scattering times could be extracted. The ICS survey searched for maximum total widths up to 12 time samples, typically in the range 10--20\,ms, while the high-time-resolution analysis resolved intrinsic widths and scattering times down to $\sim 0.01$\,ms, albeit with some ambiguity between scattering and intrinsic structures for some FRBs.

We analyse only FRBs with known redshift $z$ based
on their high-probability association to a host galaxy, giving a total sample of 29 events, and utilise their online signal-to-noise ratio S/N, and total dispersion measure \DMobs. To this we add the signal-to-noise maximising width \wsnr, and observed scattering time \tauobs, which when scaled to 1\,GHz we denote \taughz. These data, together with properties derived in this work, are reported in Table~\ref{tab:data}.

\subsection{Model of effective width}
\label{model}

We characterise the effective width \weff\ of an FRB as per \citet{Cordes_McLaughlin_2003}, using the geometric sum
\begin{eqnarray}
\weff & = & \sqrt{\wi^2 + \wtau^2 + \wdm^2 + \tres^2}, \label{eq:effwidth}
\end{eqnarray}
where the terms on the right-hand-side are intrinsic width, scattered width, DM-smearing time within each search channel, and search time resolution respectively. The experimental fluence threshold, $F_{\rm th}$, then increases as
\begin{eqnarray}
F_{\rm th}({\weff}) & = & \left( \frac{\weff}{1{\rm ms}} \right)^{0.5} F_{\rm th}(1{\rm ms}). \label{eq:Fth}
\end{eqnarray}

The exact values that should be used for each term in Eq. \ref{eq:effwidth} are not immediately obvious however. Both \tres\ and \wdm\ have a boxcar impulse response, whereas \wtau's response is an exponential, assuming a narrow bandwidth and single dominant scattering screen; and the intrinsic shape of an FRB can be very complex, and is often fit as the sum of multiple Gaussians \citep[e.g.][]{Hessels_121102_2019, qiu_population_2020,CHIME_cat1_morphology}.
Furthermore, as shown by \citet{2024MNRAS.528.1583H}, the response of FRB search algorithms (which typically use a boxcar search of varying width) can be relatively complex, especially to multi-component FRBs,
so that the effective width seen by an FRB search algorithm (as modelled by Eq.~\ref{eq:effwidth}) varies as scattering and/or DM smearing blends components together. Indeed, the use of any single value to characterise the intrinsic width of an FRB will always be an approximation, as will the use of Eq.~\ref{eq:Fth}. We observe however that only in a single FRB analysed by \citet{2024MNRAS.528.1583H} did the response curve significantly deviate from the functional form of Eq.~\ref{eq:Fth}. Therefore, we choose in this work to at least ensure that the terms used in Eq.~\ref{eq:effwidth} have appropriate weighting, so that each represents a width corresponding to $\pm1$ standard deviation of their underlying distributions. Thus we multiply \wdm\ and \tres\  by $3^{-0.5}$, corresponding to twice the standard deviation of a boxcar function compared to the total width. The standard deviation of an exponential of form $\exp(-t/\tau)$ is simply $\tau$, so we use $2 \tau_{\rm obs}$ for $w_\tau$. 

\begin{figure}
    \centering
    \includegraphics[width=0.9\linewidth]{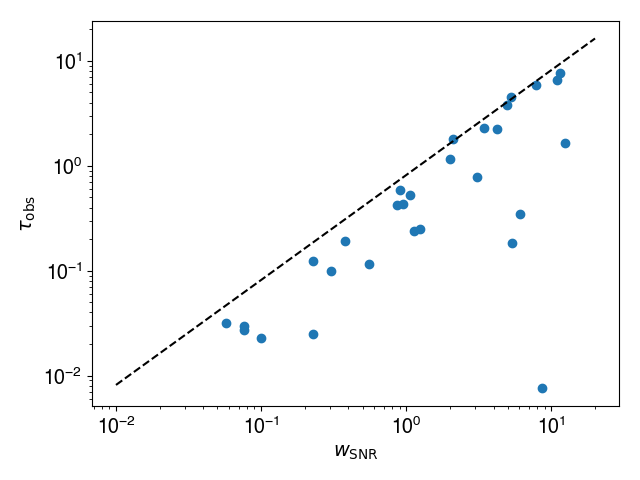}
    \caption{Plot of scattering time at central frequency, $\tauobs$, against signal-to-noise maximising width, \wsnr, for the CRAFT HTR sample with known redshift after structure-maximising dedispersion. The 1-1 line is $\wsnr = 1.225 \tauobs$. The observed correlation is consistent with observational bias, as discussed in text, and by \citet{CHIME_baseband_morphology_2025}.}
    \label{fig:snr_scat}
\end{figure}

Coherent dedispersion of the CRAFT ICS data removes both the $\tres$ and \wdm\ terms from Eq.~\ref{eq:effwidth}, so that the high-time-resolution data products robustly measure an apparent width, \wapp, given by
\begin{eqnarray}
\wapp & = & \sqrt{\wi^2 + \wtau^2} \label{eq:wapp}.
\end{eqnarray}
In this work, we set $\wapp$ to be the signal-to-noise maximising width $\wsnr$ reported by \citet{CRAFT_HTR}. For a scatter-dominated (i.e., purely exponential) signal, the S/N is maximised when $\wsnr = 1.225 \tau$. For real FRBs, with non-negligible intrinsic width, the total width will be at least as large as $1.225 \tau$
--- a constraint which is born out in the CRAFT HTR data (see Figure~\ref{fig:snr_scat}). We use this to define an intrinsic width \wi\ as
\begin{eqnarray}
\wi & = & \sqrt{\wobs^2 - (1.225 \tau)^2}. \label{eq:wi}
\end{eqnarray}
In cases where the FRB time-profile is completely dominated by scattering, the calculated value of $\wi$ becomes highly uncertain, and sensitive to the exact value of $\tau$. Indeed, for two FRBs, Eq.~\ref{eq:wi} implies $\wi^2< 0$. In these cases, we (somewhat arbitrarily) set $\wi = 0.01 \wtau$ --- this choice affects fits to the low-width part of the $\wi$ distribution. However, this work is primarily concerned with the behaviour at high time widths, which is unaffected.

The most uncertain aspects of the data from \citet{CRAFT_HTR} relate to scattering measurements, which in turn affects intrinsic width via Eq.~\ref{eq:wi}. We therefore perform a bootstrap analysis to determine how uncertainties in scattering values affect our results. To do so, we use the reported uncertainties in $\tau$ to generate Gaussian deviates, which we add the the measured values of $\tau$, producing new values of $\wi$. We then re-run our fit procedures (outlined below), repeat 100 times, and use the observed variation in fitted values as uncertainties on our results.

\subsection{Redshift dependence}
\label{sec:redshift}

We model the intrinsic width distribution in the frame of the FRB progenitor as being constant, under the assumption that there is no cosmological evolution in the small-scale physical processes that produce FRBs. Thus we assume $\wi$ in the observer frame increases as $(1+z)$ due to time-dilation.

We model FRB scattering to originate in the rest-frame of the progenitor. FRBs experience scattering from material in the vicinity of their progenitor, in their host galaxy, from intervening halos, and from the Milky Way. However, the scattering expected from intergalactic media and halos is expected to be very small \citep{macquart_temporal_2013}, with FRBs with negligible scattering being observed to pierce intervening halos \citep{ProchaskaHalo2019}, while both pulsar \citep{Bhat2004} and FRB \citep[e.g.][]{cho_spectropolarimetric_2020} observations show that scattering induced from the Milky Way at high Galactic latitudes (where our data sample is predominantly obtained) is small. \citet{2022ApJ...927...35C} performed a population synthesis of FRB scattering, finding that the host ISM alone did not explain the observed scattering distribution. Those authors noted that contributions from the circumbust medium or intervening galaxies would alleviate the tension --- however, two-screen scattering studies of FRBs have excluded the second scenario \citep{2015Natur.528..523M,sammons_two-screen_2023}, while requiring FRB scattering centres to be located in their host galaxies. The model of \citet{2022ApJ...934...71O} also finds that the scattering timescales of FRBs in the $z \lesssim 1$ Universe will mostly be host-dominated, with a small high-scattering tail due to intersections with intervening galaxies along the line-of-sight.  Therefore, the origin of significant ($\sim$ms) scattering in FRBs observed at high Galactic latitudes is expected to be set primarily in the host rest frame.

If scattering follows a power-law dependence, such that $\tau \sim \nu^\alpha$, then scattering at high redshifts will be suppressed by $(1+z)^{\alpha+1}$, where we take $z$ to be the redshift of the FRB host. We assume $\alpha=-4$ as per a Gaussian distribution of inhomogeneities in the scattering screen \citep{1975ApJ...201..532L}, although both FRBs \citep{CRAFT_HTR} and pulsars \citep{Bhat2004} show deviations from this behaviour. We do not use the values of $\alpha$ estimated by \citet{CRAFT_HTR} for the FRBs in our sample due to the high uncertainty in these values.

\subsection{Analytical form}
\label{sec:analytical form}

All models of the distribution of FRB intrinsic widths and scattering have so-far used lognormal distributions \citep{2021MNRAS.501.5319A,chimefrb_collaboration_first_2021} to describe the intrinsic distributions of these variables. For the sake of continuity, we therefore also adopt a lognormal form in both parameters. Due to both frequency-dependence in $\tau$, and redshift-dependence in both $\tau$ and $w$, we quote values of $\mu_t, \sigma_t$ in log-10 space for $t=\tau,\wi$ normalised to 1\,GHz in the rest-frame of the host (i.e., as would be viewed at $z=0$).

\begin{figure}
    \centering
    \includegraphics[width=0.9\linewidth]{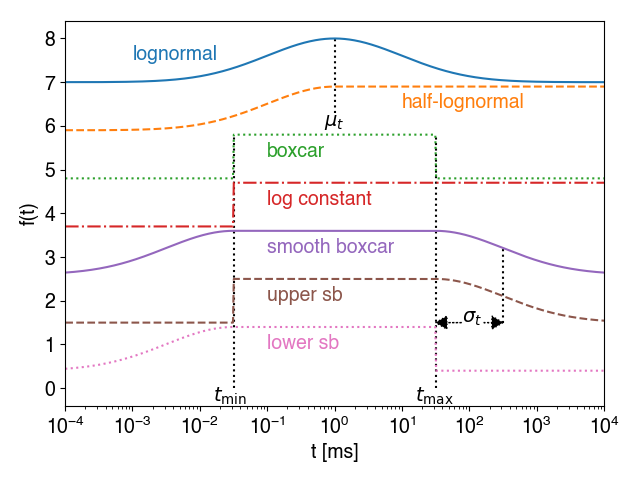}
    \caption{Illustration of the different fitting functions for the intrinsic distribution of $t=\taughz,\wi$ considered in this work; `sb' stands for `smoothed boxcar'. These functions are defined in terms of the logarithm base 10 of $t$ in ms, and parameters $t_{\rm min}$, $t_{\rm max}$, $\mu_t$, and $\sigma_t$.}
    \label{fig:functions}
\end{figure}

We also consider a variety of other models, motivated by \citet{CRAFT_HTR}, who argue that there is no evidence that the intrinsic distribution of high scattering times in either ASKAP or CHIME data is upwardly bounded by observations. We evaluate evidence for a downturn at high values by comparing the literature-standard lognormal distribution with a `half-lognormal', i.e.\ a distribution which has a Gaussian lower-half defined by $\mu_t, \sigma_t$, but is constant in log-space above the mean.

We also ensure that our conclusions are insensitive to the functional form by considering a boxcar distribution in log-space, defined by minimum and maximum values $t_{\rm min}$ and $t_{\rm max}$, and comparing this to a `log-constant' distribution defined only by a minimum value, $t_{\rm min}$, i.e.\ a step-function. We also consider boxcar distributions with smooth Gaussian downturns at the lower and/or upper edges, with are defined by the parameter set $t_{\rm min}, t_{\rm max}, \sigma_t$. A diagram showing the considered functions is given in Figure~\ref{fig:functions}.

\section{Fits to the intrinsic distributions}
\label{sec:fits}

In order to assess evidence for a downturn in scattering values, we fit functional forms from \S~\ref{sec:analytical form} to the observed intrinsic distributions of $\taughz$ and $\wi$, i.e.\ those scaled to the host frame. Each functional form is weighted by the completeness function and re-normalised to unity. The fits are performed by maximising the summed log-likelihood, ${\mathcal L} = \sum \log_{10} p(\tau)$, over all FRBs in the sample.

\subsection{Completeness calculation}
\label{sec:completeness}

Completeness functions are determined by calculating a \taumax\ and \wmax\ below which each FRB would be observable, and above which it will not be. These values can be estimated using the detected S/N values from \citet{Shannon_ICS}, a threshold S/N$_{\rm th}=10$, and noting that S/N will vary with $F_{\rm th}$ in Eq.~\ref{eq:Fth}. A key assumption of this analysis is that the distributions of $\tau$ and $\wi$ are independent, such that \taumax\ can be calculated by holding $\wi$ fixed in Eq.~\ref{eq:effwidth}, and vice versa. Thus FRBs which have intrinsically large $\tau$ necessarily probe high values of $\wi$ and vice versa, and FRBs with high S/N probe high values of both, as do those with large DM, since then DM smearing dominates $\weff$. The completeness of the ICS survey to a given $\tau$ or $\wi$ can then be defined as the fraction of all FRBs which would have been detectable had they had that value. The calculation is also repeated by scaling observed \tauobs\ and \wi, and limiting \taumax\ and \wmax, to the host rest-frame as per \S~\ref{sec:redshift}.

\begin{figure}
    \centering
    \includegraphics[width=\linewidth]{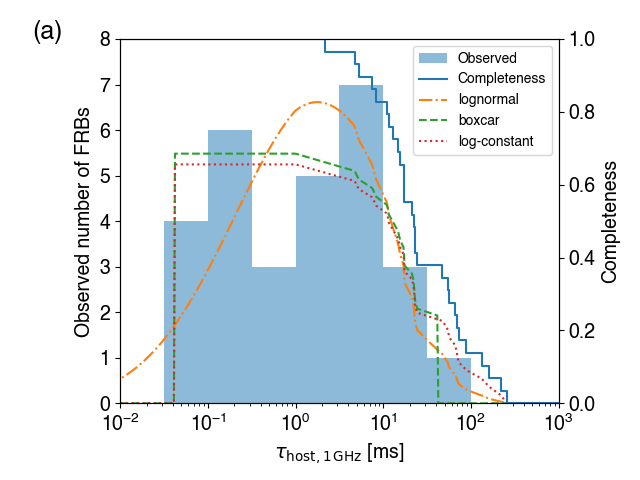}
    \includegraphics[width=\linewidth]{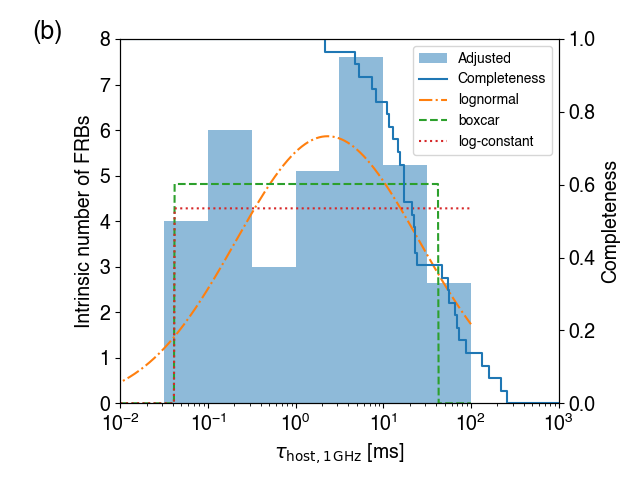}
    \caption{FRB scattering distributions. (a) The observed distribution of rest-frame scattering normalised to $1$\,GHz, $\taughz$, as well as fits to the intrinsic distribution adjusted for the completeness function; (b)
    intrinsic scattering distribution of FRBs, being the observed distribution adjusted for completeness, compared to intrinsic fitted functions.}
    \label{fig:tauhist}
\end{figure}

\begin{figure}
    \centering
    \includegraphics[width=\linewidth]{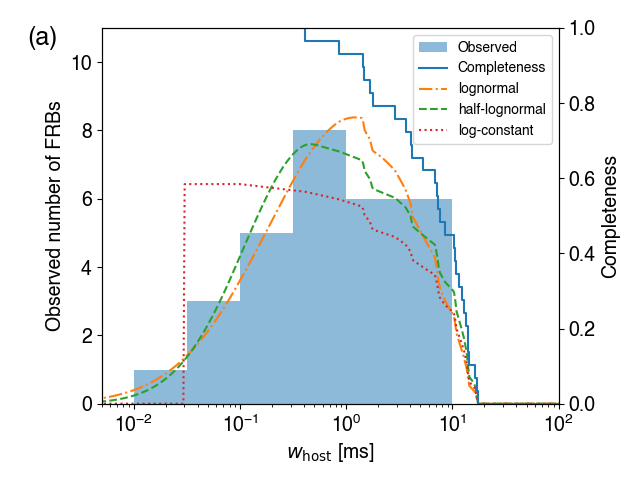}
    \includegraphics[width=\linewidth]{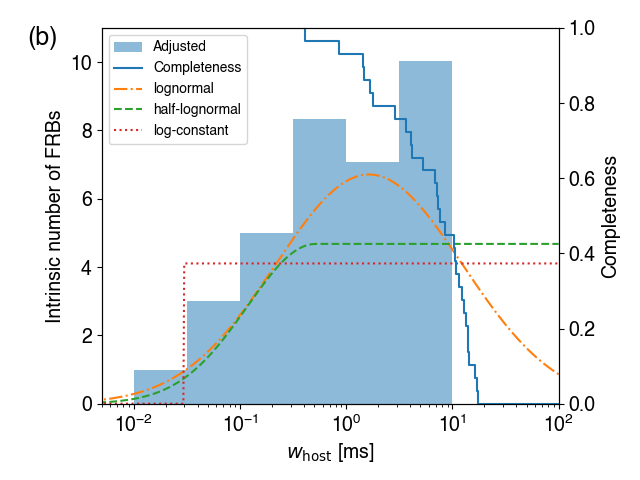}
    \caption{FRB width distributions. (a) The observed distribution of rest-frame intrinsic width, $\wi$, as well as fits to the intrinsic distribution adjusted for the completeness function; (b)
    intrinsic width distribution of FRBs, being the observed distribution adjusted for completeness, compared to intrinsic fitted functions.}
    \label{fig:whist}
\end{figure}

The resulting completeness functions for $\tau$ and $\wi$ are shown in Figures~\ref{fig:tauhist} and \ref{fig:whist}, respectively, together with histograms of observations, and corrected histograms accounting for completeness. The rapid drop in completeness in both $\tau$ and $w$ from 1--20\,ms in each plot is most evident in the observed frame; in the case of scattering, the effect of completeness is most obvious in the scaled values, $\tau_{{\rm host}, \rm 1\,GHz}$, where the observed scattering fraction closely follows the completeness above 10\,ms. The corrected histograms show no evidence for a high-scattering/width downturn, but rather cut off abruptly close to the points where completeness drops to zero.

\subsection{Fit results}
\label{sec:fit_results}

\begin{table*}[]
    \centering
    \begin{tabular}{l|l l l l}
    Name & Parameters & Values & $\log_{10} {\mathcal L}_{\rm max}$ & $ \log_{10} {\mathcal L}_{\rm max} - \log_{10}{\mathcal L}_{\rm max}^{\rm lognormal}$ \\
    \hline
lognormal & $\mu_\tau, \sigma_\tau  $ & $0.36 \pm 0.19, 1.05 \pm 0.08 $ & -14.29 & N/A \\
half-lognormal & $\mu_\tau, \sigma_\tau  $ & $-1.38 \pm 0.68, -0.00 \pm 0.47 $ & -12.80 & $1.49 \pm 0.58 \, (0.66)$\\
boxcar & $\tau_{\rm min},\tau_{\rm max}  $ & $-1.38 \pm 0.28, 1.63 \pm 0.19 $ & -12.25 & $2.04 \pm 0.52 \, (0.77)$ \\
log-constant & $\tau_{\rm min}  $ & $-1.38 \pm 0.28 $ & -12.80 & $1.49 \pm 0.49 \, (0.66)$ \\
smooth boxcar & $\tau_{\rm min},\tau_{\rm max},\sigma_\tau  $ & $-1.39 \pm 0.42, 2.97 \pm0.89, -0.00 \pm 0.30 $ & -12.80 &  $1.49 \pm 0.58 \, (0.66)$\\
upper sb & $\tau_{\rm min},\tau_{\rm max},\sigma_\tau  $ & $-1.39 \pm 0.18, 1.62 \pm 0.41, -0.00 \pm 0.65 $ & -12.25 &$2.04 \pm  0.53 \, (0.72)$\\
lower sb & $\tau_{\rm min},\tau_{\rm max},\sigma_\tau  $ & $-1.39 \pm 0.43, 2.97 \pm 0.83, -0.00 \pm 0.30 $ & -12.80 & $1.49 \pm 0.58 \, (0.66)$\\
    \end{tabular}
    \caption{Best-fitting parameters for different functions fit to $\log_{10} \tauhost {\rm [ms]}$, i.e.\ the intrinsic distribution of scattering times at 1\,GHz, \tauhost. Also shown is the difference in maximum likelihood with respect to the lognormal model, with uncertainties given by bootstrapping of scattering only, with the uncertainty when including scattering index shown in brackets.}
    \label{tab:scat_results}
    \medskip
\raggedright $^a$ The likelihood for these functions is almost flat in the range $\taumax \ge 1.62 $ region, and returned fit values show numerical fluctuations between 1.62 and 2.97, depending on the exact initial guesses.  Uncertainties represent variation from our bootstrap results (see text).
\end{table*}

\begin{table*}[]
    \centering
    \begin{tabular}{l|l l l l}
    Name & Parameters & Values & $\log_{10} {\mathcal L}_{\rm max}$  & $ \log_{10} {\mathcal L}_{\rm max} - \log_{10}{\mathcal L}_{\rm max}^{\rm lognormal}$ \\
    \hline
lognormal & $\mu_w, \sigma_w  $ & $0.22 \pm 0.04, 0.88 \pm 0.03 $ & -10.48 & N/A \\
half-lognormal & $\mu_w, \sigma_w  $ & $-0.29 \pm 0.14, 0.65 \pm 0.08 $ & -10.37 & $0.11 \pm 0.05 \, (0.05)$  \\
boxcar & $w_{\rm min},w_{\rm max}  $ & $-1.53 \pm 0.11, 2.61 \pm 0.04 $ & -10.24 & $0.24 \pm 0.37 \, (0.5)$ \\
log-constant & $w_{\rm min}  $ & $-1.53 \pm 0.11 $ & -10.24 & $0.24 \pm 0.37 \, (0.5)$\\
smooth boxcar & $w_{\rm min},w_{\rm max},\sigma_w  $ & $-0.29\pm0.11, 2.14 \pm 0.55, 0.65 \pm 0.06 $ & -10.37 & $0.11 \pm 0.05 \, (0.05)$ \\
upper sb & $w_{\rm min},w_{\rm max},\sigma_w  $ & $-1.52 \pm 0.11, 2.31 \pm 0.02, 1.08 \pm 0.01 $ & -10.18 & $0.30 \pm 0.37 \, (0.51)$ \\
lower sb & $w_{\rm min},w_{\rm max},\sigma_w  $ & $-0.29 \pm 0.14, 2.93 \pm 0.31, 0.65 \pm 0.08 $ & -10.37 & $0.11 \pm 0.05 \, (0.05)$ \\
    \end{tabular}
    \caption{Best-fitting parameters for different functions fit the to distribution of $\log_{10} \whost {\rm [ms]}$, i.e.\ the intrinsic host width. Uncertainties represent variation from our bootstrap results (see text).}
    \label{tab:width_results}
\end{table*}

Using the above-mentioned completeness functions, resulting best-fit functions for $\taughz$ and $\wi$ are compared to the observed distributions in Figures~\ref{fig:tauhist} and Figures~\ref{fig:whist}, respectively. Numerical values of best-fits are given in Tables~\ref{tab:scat_results} and \ref{tab:width_results}, respectively. These tables also show our estimated uncertainties in relative likelihood differences between models, which we estimate using a bootstrap resampling technique for both $\tau_{\rm obs}$ and $\alpha$, described in \S~\ref{sec:robustness}.

In the case of scattering, we find that fits to the half-lognormal, and to the smoothed edge boxcar functions, prefer infinitesimal standard deviations, i.e.\ they tend to distributions with sharp edges, which are identical to the log-constant and boxcar distributions, respectively. The best-fitting lognormal ($\mu_\tau,\sigma_\tau=0.36, 1.05$) provides a  worse likelihood, with $\log_{10} {\mathcal L}=-14.29$, compared to the boxcar ($\tau_{\rm min},\tau_{\rm max}=-1.38,1.63$; $\log_{10} {\mathcal L}=-12.25$) and log-constant distributions ($\tau_{\rm min} = -1.38$; $\log_{10} {\mathcal L}=-12.8$), with significance of approximately $3 \sigma$, according to our bootstrapped error estimation.

We suggest therefore that a better fit is a log-uniform distribution over the range of at least 0.04--40\,ms, which may extend to slightly lower, but in particular much higher, values.

For intrinsic width $\wi$, no fits favour a high-width cutoff, so that the boxcar, boxcar with high-width smooth downturn, and log-constant distributions, all produce equivalent fits ($\log_{10} \mathcal{L}=-10.21 \pm 0.03$, $w_{\rm min}=-1.53$) --- we characterise these with the function with the least parameters, i.e.\ a boxcar. Similarly, the boxcar with smooth downturns at both edges, boxcar with a low-width downturn, and the half-lognormal, also produce equivalent fits ($\log_{10} \mathcal{L}=-10.37$, $w_{\rm min}/\mu_w = -0.29$, $\sigma_w=0.65$), with the half-lognormal being the function with the least parameters. A lognormal produces a distinct fit ($\log_{10} \mathcal{L}=-10.48$, $\mu_w=0.22$, $\sigma_w=0.88$). These models, and their completeness-adjusted fits, are shown in Figure~\ref{fig:whist}(b). Compared to fluctuations in likelihood differences estimated from our bootstrapping procedure, we find that alternative models show at-best a $2 \sigma$ preference beyond the lognormal distribution, which we do not consider to be significant. Thus, while we conclude again that there is no evidence for a high-width downturn in the intrinsic distribution of FRB widths, we cannot exclude a lognormal distribution as being the true underlying model. Within our alternative models, our preferred scenario nonetheless suggest that the true distribution either rises as a Gaussian in the range 0.03--0.3\,ms, and is most likely log-uniform above this value, or else is log-uniform over the entire range from 0.03\,ms upwards.

\subsection{Robustness to uncertainties}
\label{sec:robustness}

\begin{figure}
    \centering
    \includegraphics[width=\linewidth]{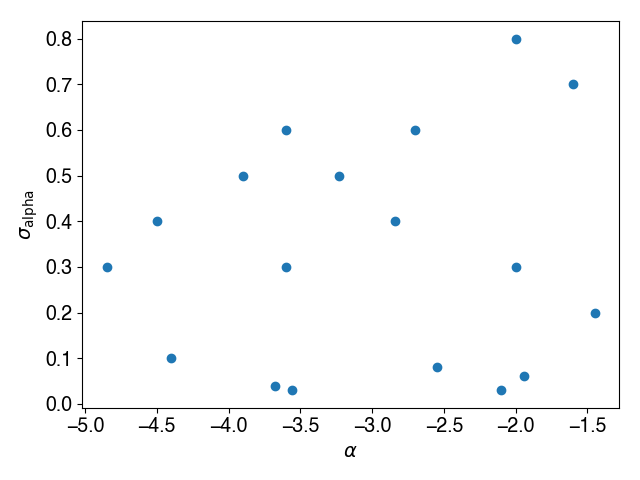}
    \caption{Values of scattering index $\alpha$, and its estimated error $\sigma_\alpha$, from the FRB sample of \citet{CRAFT_HTR}. Only those FRBs with well-measured scattering are shown.}
    \label{fig:good_scattering}
\end{figure}

Our fitting results rely on assumed values of the scattering index $\alpha$, and measured values of the scattering time $\tau$ from \citet{CRAFT_HTR}. Within the FRB literature, assumptions of $\alpha=-4$ or $\alpha=-4.4$ are common, e.g.\ the CHIME scattering measurements were scaled to 600\,MHz using $\alpha=-4$ \citep{chimefrb_collaboration_first_2021}. This is based on thin-screen scattering theory, where Gaussian and Kolmorgorov distributions of turbulence predict $\alpha=-4$ and $\alpha=-4.4$, respectively \citep{1971ApJ...164..249L}. However, there is increasing evidence that many pulsars follow a flatter dependence, with $\alpha > -4$ \citep[see e.g.\ ][]{2019ApJ...874..179K}, which might be attributable to a minimum spatial turbulence scale \citep{2009MNRAS.395.1391R}. Of the 15 FRBs in \citet{CRAFT_HTR} with well-measured scattering values ($\sigma_\alpha < 1$,$\sigma_\tau/\tau < 0.1$), the majority of scattering indices are very close to, but larger than, $-4$ --- see Figure~\ref{fig:good_scattering}.

Given the extreme nature of the plasmas in the immediate vicinity of an FRB progenitor \citep{Hessels_121102_2019}, the potential of surrounding structures such as persistent radio sources for which there is no Galactic analogue \citep[PRS; ][]{2022ApJ...927...55L} to induce scattering, the existing evidence from the Pulsar literature, and the measurements from \citet{CRAFT_HTR}, it is pertinent to consider the applicability of our assumption on $\alpha$.

\begin{figure}
    \centering
    \includegraphics[width=\linewidth]{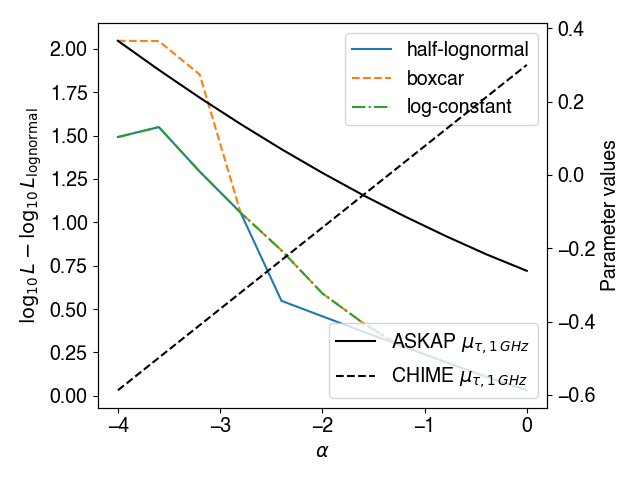}
    \caption{Difference in log-likelihoods in scattering fits between alternative models and a lognormal fit, as a function of scattering index $\alpha$.}
    \label{fig:varalpha}
\end{figure}

We consider two cases. Firstly, we vary $\alpha$ uniformly between $0$ and $-4$. This covers both the the central range of $-4 < \alpha < -2$ spanned by most of the well-measured values, and it allows for FRB scattering to scale more weakly than $\taughz \sim (1+z)^{3}$, e.g., if turbulence is not dominated by the host galaxy.
The results for scattering fits are shown in Figure~\ref{fig:varalpha} --- there is no effect on the width distribution. For values of $\alpha < -3$, alternatives to the lognormal model are significantly preferred, while as $\alpha$ increases above $-3$, differences in preferences between functional forms drops markedly. From Figure~\ref{fig:good_scattering},  all well-measured values of $\alpha$ have $\alpha < -1.5$, so we do not consider it plausible that the true mean of the $\alpha$ distribution lies above $-3$. Therefore, we consider our result that a distribution without a high-scattering downturn is preferred to a lognormal distribution to be robust.

As $\alpha$ increases, the mean of the lognormal distribution $\mu_\tau$, fitted to ASKAP data, decreases, while the implied value at 1\,GHz of CHIME fits at 600\,MHz increases, such that the two are consistent at $\alpha=-1.7$ (there is little effect on $\sigma_\tau$, which remains discrepant). Combined with the inability of \citet{chimefrb_collaboration_first_2021} to correct for redshift, this may well explain differences between our results and theirs.

Our second case uses bootstrapping to vary both the measured values of $\tau$ and $\alpha$ within their error ranges, and applies the observed value of $\alpha$ to each FRB individually. We sample each error from a normal distribution with standard deviation equal to the estimated error, and re-draw from the distribution in cases where a negative value of $\tau$ is produced. In cases where the error on $\alpha$ is very large, however, this procedure can produce unrealistic values. Hence, when $\sigma_\alpha > 1$, we instead sample $\alpha$ from a normal distribution with mean $-3$ and standard deviation of unity, which approximately mimics the distribution of well-measured points in Figure~\ref{fig:good_scattering}. We have reported the mean, and root-mean-square, differences in log-likelihood values between the lognormal and alternative models in Tables~\ref{tab:scat_results} and \ref{tab:width_results}, both with resampling $\tau$ only, and also with resampling $\alpha$. For distributions of scattering, we find typical uncertainty in log-likelihood differences of 0.5 (0.7) excluding (including) variation in $\alpha$, leading to our conclusion that if $\alpha=-4$, models excluding a downturn are strongly favoured, while there is a relatively small preference at $\alpha=-2$ and higher.

\section{Influence on FRB population models}
\label{sec:zdm}

Models of the FRB population, accounting for the dispersion measure budget \citep[e.g.][]{Macquart2020} and FRB population and luminosity function \citep[e.g.][]{2018MNRAS.481.2320L}, are subject to biases due to the selection function of the detecting instrument, which, if not folded into the analysis, may lead to inaccurate results \citep{Connor2019}. Several works have therefore included models of the FRB intrinsic width and scattering distributions when modelling FRB data \citep[e.g.\ ][]{james_zdm_2022,2023ApJ...944..105S}, while CHIME have developed a pulse injection system \citep{2023AJ....165..152M} to model the instrumental response \citep{chimefrb_collaboration_first_2021}. These distributions are important because they modulate the effects of DM smearing: intrinsically narrow FRBs have their burst width increased significantly by DM smearing, making them less likely to be detected at high DM, whereas broad FRBs --- due to either scattering or intrinsic width --- are less affected. A potential implication of a large population of highly scattered FRBs therefore is that at high redshifts, the $(1+z)^{-3}$ suppression of scattering time will result in a higher detection rate, which, if not modelled correctly, might be mistaken for evolution in the underlying FRB population.

We therefore implement our model of scattering into the {\sc zDM} code, and test for such a correlation.

\subsection{Implementation in {\sc zDM}}
\label{sec:implemetation}

The \zdm\ code was developed to model FRB observations, including experimental biases, and allow FRB data to constrain both the intrinsic FRB population and cosmological parameters \citep{james_zdm_2022}. The latest iteration is described in \citet{2025PASA...42...17H}, which uses a Markov-Chain Monte Carlo
(MCMC) implemented by the Python {\sc emcee} package \citep{emcee}, with the likelihood calculated as
\begin{eqnarray}
{\mathcal L} & = & p_n(N_{\rm FRB}) \, \Pi_{i=1}^{N_{\rm FRB}} p_{s}(s_i|z_i,{\rm DM}_i) p_{\rm zdm} (z_i,{\rm DM}_i), \label{eq:likelihood}
\end{eqnarray}
where the $p_n$ is the probability of a survey observing $N_{\rm FRB}$ FRBs, $p_{\rm zdm}$ is the probability of FRB $i$ having redshift $z_i$ and extragalactic dispersion measure DM$_i$, and $p_s$ is the probability of that FRB having relative signal-to-noise ratio ${\rm S/N}_{i} = s_i {\rm S/N}_{\rm th}$ compared to the detection threshold ${\rm S/N}_{\rm th}$.

Internally, the \zdm\ code integrates over possible values of FRB apparent width \wapp\ (defined by Eq.~\ref{eq:wapp}) and beam sensitivity $B$, using the sensitivity model of Eq.~\ref{eq:Fth} and effective width given by Eq.~\ref{eq:effwidth}, which adds DM smearing and instrumental time resolution to \wapp. To constrain width and scattering distributions, we therefore extend Eq.~\ref{eq:likelihood} with the probability of observing \wapp\ given other parameters, $p_w(\wapp|s,z,DM)$, and probability $p_\tau(\taughz|\wapp,z)$ of observing scattering time $\taughz$ given the FRB originates from redshift $z_i$ and has apparent width $\wapp$. Note that, given $\taughz$ and $\wapp$, the probability of observing a given intrinsic width $p_{wi}(\wi| \taughz,\wapp)$ is analytically unity from Eq.~\ref{eq:wapp}.

We describe distributions $p(\weff)$, $p(\taughz)$, $p(\wi)$ using histograms with 33, 100, and 100 bins respectively, between 0.01\,ms and 1000\,ms. The coarser binning in $p(\weff)$ reflects the slow variation of sensitivity with $\weff$ and the calculation time required to integrate $p_{zdm}$ over $\weff$, while the finer binning in $p_\tau$ and $p_{wi}$ reflects our more precise ability to measure these parameters offline. The coarse $\weff$ binning however means that $\wi$ is not completely determined by $\weff$ and $\tau$. We therefore modify our calculation of $p_\tau$ to reduce numerical approximations, as $p_\tau = 0.5 p(\taughz|\weff,z) + 0.5 p(\wi|\weff,z)$.

Currently, there is no method to incorporate the tendency for repeating FRBs to have broader widths and higher scattering than apparent non-repeaters, as found by \citet{CHIME_cat1_morphology}, nor incorporate any redshift evolution in these parameters (other than the redshift dependencies described in \S~\ref{sec:redshift}).

\subsection{Effect of scattering on FRB rates}
\label{sec:zrates}

\begin{figure}
    \centering
    \includegraphics[width=\linewidth]{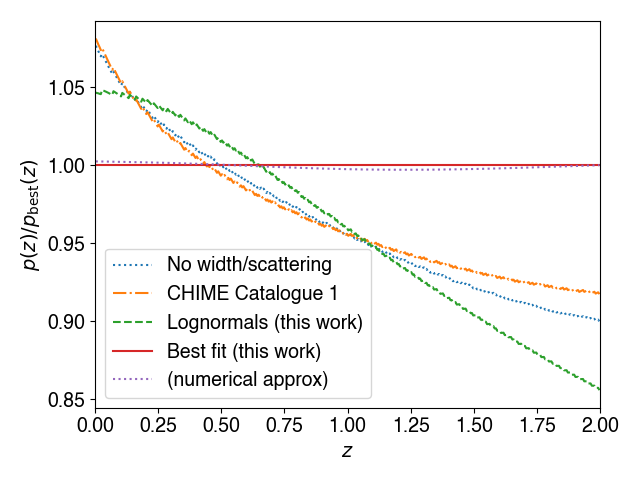}
    \caption{Relative FRB detection rate as a function of redshift for the ASKAP/CRAFT ICS survey at 1.3\,GHz, relative to the best-fit distributions from this work, for different models of FRB scattering and width (see text).}
    \label{fig:relz}
\end{figure}

To illustrate the degree of importance of modelling FRB width and scattering, we use the CRAFT ICS survey data for 1.3\,GHz observations described by \citet{Shannon_ICS} and modelled by \citet{2025PASA...42...17H}. We then use five methods to describe the scattering-width distribution. Firstly, we ignore it, and treat all FRBs as having an apparent width \wapp\ of 1\,ms. Secondly, we use the fiducial model of \citet{chimefrb_collaboration_first_2021}, which models FRB width and scattering as lognormal distributions, but includes no redshift dependence. Thirdly, we use updated lognormal fits from Tables~\ref{tab:scat_results} and \ref{tab:width_results}, including redshift scaling from \S\,\ref{sec:redshift}. Fourthly, we use what we consider to be our most likely models --- a log-constant distribution in scattering, and a half-lognormal in width, again from Tables~\ref{tab:scat_results} and \ref{tab:width_results}. These last three distributions are modelled as discrete histograms in width and scattering with 100 bins each. Lastly, we use these best-fits, but use only five histogram bins, which is the standard in \zdm. Note that all of these models have a DM-dependent FRB effective width, and hence detection threshold, due to the DM-smearing term $w_{\rm DM}$ in Eq.~\ref{eq:effwidth} --- the difference is in how the terms $\wi$ and $\wtau$ are treated.

The resulting relative distributions of FRBs in redshift are given in Figure~\ref{fig:relz}. Alternative models show an excess of $\sim$5\% at $z=0$, a deficit of $\sim5$\% at $z=1$, and $\sim 10$\% at $z=2$, compared to our most likely model. This behaviour is expected, due to the effect of redshift shifting the large number of highly scattered FRBs in the host rest-frame to lower observed scattering times, making these FRBs more detectable. Importantly, the effect of numerical approximations using only five bins for total FRB width is less than 1\%, showing that width/scattering can be accurately modelled in good computational time. By predicting 15\% more FRBs at $z=1$ relative to $z=0$ compared to alternative models, estimates of FRB population parameters will need to adjust to compensate when fitting experimental data. Thus we expect that future fits to data will predict less source evolution, a steeper spectral index, and/or a steeper (negative) FRB spectral dependence due to these updates. We remind readers however that full fits to the FRB population are model-dependent and involve several correlated parameters \citep{2026PASA...43...17H}, and the effects of FRB width and scattering also depend on experimental parameters, so that the effects discussed above may not hold quantitatively.

\subsection{Parameter estimation}
\label{sec:zdmresults}

To determine the ability of {\zdm} to constrain FRB width and scattering distributions, we use a half-lognormal function, since this reflects our lack of evidence for a high-width or scattering cutoff, but allows the shape of the distribution at low values to be modelled. We also allow the parameter $\nsfr$ to vary, which sets FRB population evolution to scale with the star-formation rate to the power $\nsfr$, to search for any correlations of these fits with the star-formation rate. We fix all other parameters --- which describe the FRB luminosity function, and host galaxy DM contributions --- to the best-fit values found by \citet{2025PASA...42...17H}, with the Hubble Constant constrained to the known range (approximately 70\,km\,s$^{-1}$\,Mpc$^{-1}$). Thus our fits are to $\mu_w,\sigma_w$, $\mu_\tau,\sigma_\tau$, and $\nsfr$.

\begin{figure*}
    \centering
    \includegraphics[width=0.75\linewidth]{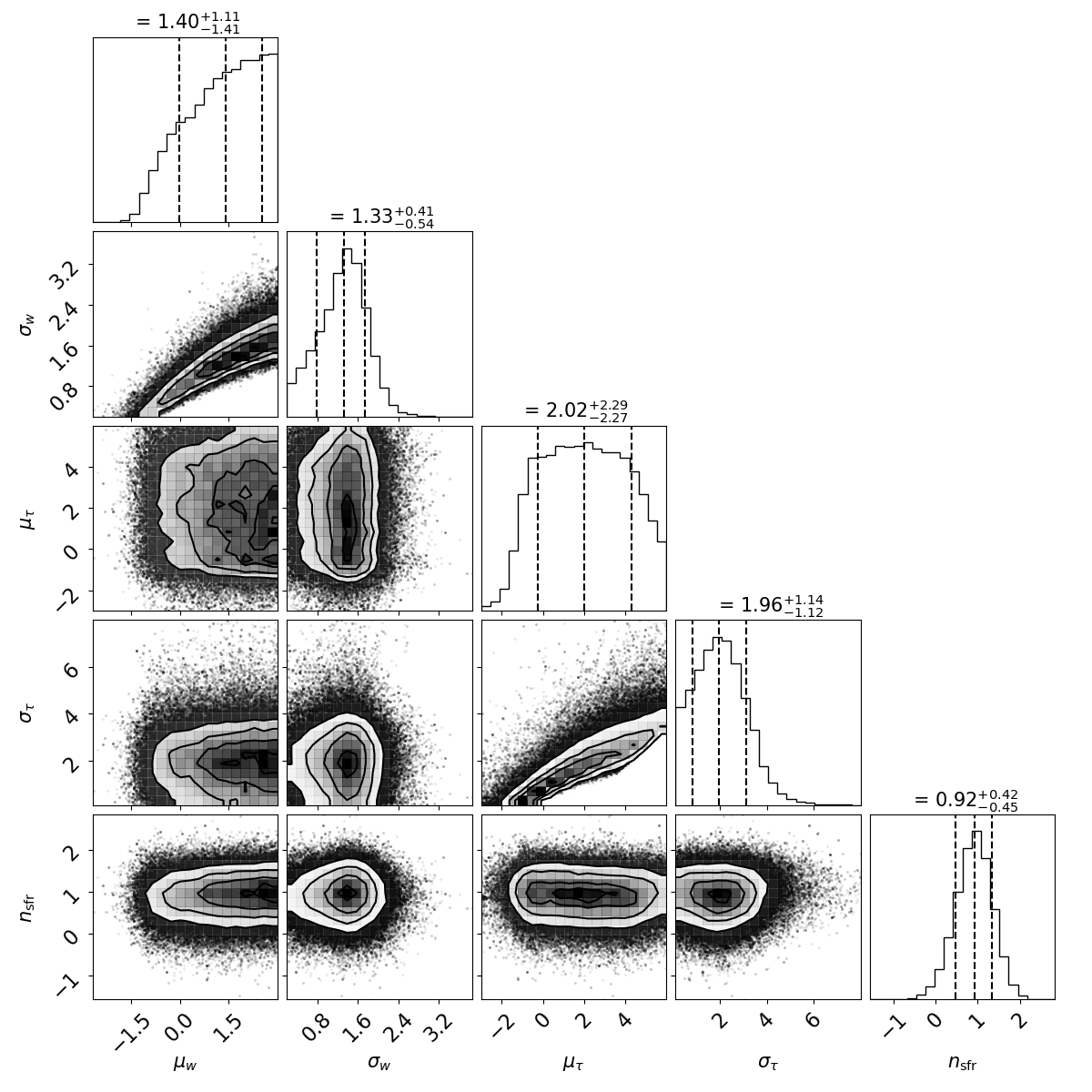}
    \caption{Corner plot of MCMC results when fitting parameters $\mu_w,\sigma_w$, $\mu_\tau,\sigma_\tau$, and $\nsfr$ to the CRAFT ICS HTR data in \zdm.}
    \label{fig:cornerplot}
\end{figure*}

The CRAFT/ICS sample, for which HTR data is available, has already been implemented in \zdm\ --- we simply add width and scattering data from \citet{CRAFT_HTR}. We maximise the likelihood described in \S~\ref{sec:implemetation}, and run the MCMC using 80 walkers with 4000 iterations per walker, finding a typical burn-in time of 100 steps (we discard the first 200 out of caution). The resulting corner plot is shown in Figure~\ref{fig:cornerplot}.

We find fits which are broadly consistent with the maximum likelihood fits of  \S~\ref{sec:fit_results}, but which highlight the relevant correlations in the fits. For $\taughz$, a most likely value at $\mu_\tau,\sigma_\tau \sim -1, 0$ is found, but there is a broad degeneracy in likelihood where both $\mu_\tau$ and $\sigma_\tau$ are positively correlated. The fits for $\wi$ show similar behaviour, with a strong degeneracy between $\mu_w$ and $\sigma_w$, for exactly the same reasons, although the most likely values are higher. Both highlight that we do not observe any measurable peak in the $\taughz$ or $\wi$ distributions, and that the observed values are dominated by selection effects, making it difficult to uniquely constrain the intrinsic distributions.

Most importantly, we find no evidence for correlation with FRB source evolution, as given by $\nsfr$. Note that this is not inconsistent with our findings from \S\,\ref{sec:zrates}, since that investigated the effects of the modelling method, rather than parameter degeneracies within that method. In effect, our lack of correlation with $\nsfr$ states that our width and scattering data sufficiently constrains the redshift-dependent bias exhibited in Figure~\ref{fig:relz}, and that this is not affected by the parameter degeneracies between $\mu_\tau$ and $\sigma_\tau$, and $\mu_w$ and $\sigma_w$.

Note that we do not consider the 1D projections of posterior probabilities for individual parameters to be meaningful, both because several parameters push against the boundaries set by priors in log-space, and because we do not consider these priors to be well-motivated.

\section{Discussion}
\label{sec:discussion}

\subsection{Comparison with other results}

In this work, we have found that the intrinsic FRB scattering and width distributions show no evidence for downturns, but may continue to values much higher than the $\sim$0.01--10\,ms range currently probed by experimental data. This is consistent with the discussion presented in \citet{chimefrb_collaboration_first_2021}, who note that their fiducial model of a lognormal distribution in scattering is poorly constrained at high scattering times. Assuming a frequency dependence of scattering time $\tau \sim \nu^{-4}$, we exclude the quantitative values of their fiducial model, which --- when expressed in $\log_{10}$, and scaled to 1\,GHz --- are $\mu_\tau,\sigma_\tau = -0.58,0.75$ and $\mu_w,\sigma_w = 0,0.42$, with these values being less likely by a factor of $10^{-7}$ for \taughz, and $10^{-16}$ for \wi, compared to our most likely values.

The cause of this difference could be intrinsic --- e.g., the true dependence of FRB scattering with frequency may not scale as $\tau \sim \nu^{-4}$. Varying $\alpha$ above $-4$ shows that our best fits have only a ten times higher likelihood than CHIME's scattering fits for $\alpha=-2$, although the width distribution is strongly ruled out in all cases. Alternatively, FRBs may have frequency-dependent widths. Artificial differences could also be the cause, e.g.\ CHIME's fits were performed to observational data, and were not scaled to the host rest frame since most hosts were unknown; their fits to bias in width and scattering were uncorrelated, and thus could not account for correlations in the bias; and these fits were performed with low-time-resolution, low-S/N data --- \citet{CHIME_baseband_morphology_2025} have re-fit the scattering times and intrinsic widths to a sub-sample of 137 FRBs for which baseband data was available, but did not estimate the intrinsic distributions of these parameters.

Several other works have probed FRB distributions in the high-width, high scattering time regime. \citet{2022MNRAS.515.3698C} report four FRBs found in Murriyang (Parkes) data, with total widths ranging from 50--200\,ms; however, two of these FRBs have very low S/N ratios, while another has a spectrum which is very similar to temporally coincident RFI, so that we consider only one of these --- FRB~19910730A, with width 113\,ms --- to be a firm detection. The CRAFT coherent upgrade, CRACO, has reported two FRBs detected at an integration time of 110\,ms \citep{2025PASA...42....5W}, although no high-time-resolution data was available to differentiate between scattering and intrinsic width. Perhaps the best evidence for a large number of intrinsically high-width, high-scattering FRBs comes from studies of high-redshift FRBs, which probe high intrinsic scattering values due to the $(1+z)^3$ suppression in the observer frame. \citet{2025arXiv250801648C} report FRB~20240304B with modest $\taughz=5.3 \pm 0.3$\,ms; its high redshift of $z=2.148$ implies a rest-frame $\taughz=165$\,ms. \citet{2025ApJ...993..208S} discuss FRB~20200723B, which has a scattering time $\tau_{\rm 600\,MHz} = 340 \pm 10$\,ms, scaling to 44\,ms at 1\,GHz (its likely host, NGC 4602, resides at negligible redshift).

It is difficult to use the above works to quantitatively probe the FRB intrinsic width and scattering distributions, however, since they are not part of FRB surveys publishing complete results for a large data sample. Thus we use these cases as anecdotal evidence only.

\subsection{FRB host galaxy properties}

Analysis of FRB host galaxies is often used to identify their progenitors, by comparing e.g.\ the FRB offset from host centre, and host size, metallicity, and star-formation rate, to that of other transient phenomena \citep[e.g.][]{Bhandari+22,gordon_demographics_2023, 2024Natur.635...61S}; and to search for differences between repeating, and apparently non-repeating, FRBs \citep{2025arXiv250620774M}. FRB scattering should influence these properties, with larger, more gas-rich galaxies inducing more scattering, while FRBs occurring on the outskirts of a galaxy, and/or propagating perpendicular to the  plane of the galaxy, should show less scattering. Such an effect has been reported by \citet{2024Natur.634.1065B}, who find a deficit of FRBs from galaxies with large inclination angles, and suggest scattering selection bias as the culprit.

By showing that models with no high-value downturn give an equivalently good fit to the intrinsic FRB width distribution, and a significantly improved fit to the FRB scattering distribution, we infer that the total FRB rate must be significantly limited by observational completeness, and that such an observational bias must exist at some level. However, its effect on observed host galaxy properties will depend on the intrinsic distribution of FRBs within the host galaxy, and the nature of the scattering medium. \citet{2020ApJ...903..152H} have found that FRBs have significant offsets from their host centres, but cautioned that scattering bias would distort the observed distribution to larger radii than the true distribution, i.e.\ this observation may be proof of the existence of such a bias. \citet{2025PASA...42..157G} however have searched for an association between apparent host galaxy offset and FRB scattering timescale, and found no evidence for a correlation, although whether such a correlation would have been observable given the data is unknown. Such a lack of a correlation suggests that either at most a weak bias exists, or that it is strong enough to act on FRBs even far outside the half-light radius of the host. Scattering bias may also obscure potential associations with structures within a host, for instance suppressing the fraction of FRBs associated with star-forming regions such as spiral arms, as probed by \citet{2025ApJ...993..119G}.

One reason for no bias existing would be if scattering is dominated by turbulent media associated with the FRB progenitor itself, and would thus be independent of location with the host, i.e., it behaves like the intrinsic width bias. While constraints on the host scattering medium are consistent with an origin with the host galaxy interstellar medium \citep{2015Natur.528..523M,sammons_two-screen_2023,2025Natur.637...48N}, \citet{2022ApJ...927...35C} simulate FRB scattering by placing FRBs in spiral, dwarf, and elliptical hosts, and find that the resulting scattering distribution when accounting for host and Milky Way ISMs alone cannot explain the large number of high-scattering FRBs seen by \citet{chimefrb_collaboration_first_2021} --- although those conclusions will be model-dependent.

An alternative might be the contribution of intervening halos. \citet{2022ApJ...934...71O} model FRB scattering from the Milky Way, intervening line-of-sight galaxies, and FRB hosts, finding that the relative importance of different sources of scattering depends on FRB redshift, and specific local Universe and Galactic structures. For redshifts in the range of $0.5 \le z \le 1$, FRB host galaxies are expected to dominate, but the distribution should peak below 1\,ms. However, the simulated (and subdominant) contribution from intervening halos produces a relatively flat contribution extending up to a second and beyond at 1\,GHz, and the authors note that their predictions may be conservative if turbulence scales extend below 1\,a.u. 

We can therefore offer no resolution to the above tensions, other than to emphasise that FRB scattering absolutely is a significant biasing factor in detecting FRBs, and hence, in interpreting host galaxy data; and that we trust that an updated model of the intrinsic FRB scattering distribution will help to resolve such tensions in the future.

\section{Conclusion}

We have analysed the intrinsic scattering and width distributions from 29 localised FRBs detected by the CRAFT survey science project on ASKAP. Accounting for completeness, we find that the distributions show no evidence for a downturn in the range up to $\sim10$\,ms probed by the data. The intrinsic scattering distribution is consistent with a log-uniform distribution in the 0.04--40\,ms range, and a lognormal form is a factor of $\sim 100$ less likely; while all fits to the intrinsic width distribution give similarly good results over the 0.03--30\,ms range. We caution against assuming lognormal distributions for either distribution, since these predict high-time downturns in regions currently unprobed by experimental data.

Our findings show that FRB observations are fundamentally limited by their ability to probe high values of scattering and width, and that this should produce biasing effects on host galaxy studies --- though whether or not this bias is evident in current data is disputed. The lack of an obvious bias in data might be explained by a significant fraction of FRB scattering occurring in the progenitor.

When implemented in the {\sc zDM} code, our updated width and scattering models increase the expected number of FRBs detected by ASKAP at $z=1$ compared to those at $z=0$ by $\sim$10\%. This illustrates how such `nuisance parameters' can affect FRB population estimates. We do not find any evidence of correlation between estimates of the FRB width/scattering distribution and FRB population evolution.

\begin{acknowledgement}
This scientific work uses data obtained from Inyarrimanha Ilgari Bundara, the CSIRO Murchison Radio-astronomy Observatory. We acknowledge the Wajarri Yamaji People as the Traditional Owners and native title holders of the Observatory site. CSIRO’s ASKAP radio telescope is part of the Australia Telescope National Facility (https://ror.org/05qajvd42). Operation of ASKAP is funded by the Australian Government with support from the National Collaborative Research Infrastructure Strategy. ASKAP uses the resources of the Pawsey Supercomputing Research Centre. Establishment of ASKAP, Inyarrimanha Ilgari Bundara, the CSIRO Murchison Radio-astronomy Observatory and the Pawsey Supercomputing Research Centre are initiatives of the Australian Government, with support from the Government of Western Australia and the Science and Industry Endowment Fund.

This work was performed on the OzSTAR national facility at Swinburne University of Technology. The OzSTAR program receives funding in part from the Astronomy National Collaborative Research Infrastructure Strategy (NCRIS) allocation provided by the Australian Government, and from the Victorian Higher Education State Investment Fund (VHESIF) provided by the Victorian Government.

\end{acknowledgement}

\paragraph{Funding Statement}

MG is supported by the UK STFC Grant ST/Y001117/1. MG acknowledges support from the Inter-University Institute for Data Intensive Astronomy (IDIA). For the purpose of open access, the author has applied a Creative Commons Attribution (CC BY) licence to any Author Accepted Manuscript version arising from this submission.

\paragraph{Competing Interests}

The authors declare no competing interests.

\paragraph{Data Availability Statement}

Data from this work is available within the {\sc zDM} GitHub repository, \url{https://github.com/FRBs/zdm/}.

\printendnotes

\printbibliography

\begin{table*}[]
    \centering
    \begin{tabular}{l|cccccccccccccc}
FRB & \DMobs\ & $z$ & S/N & $\nu$ & \tres\ & \wsnr\ & \tauobs\ & \taumax & \tauhost\ & \htaumax\ & \wi & \wmax\ & \whost\ & \hwmax\ \\
& [\pccc] & & & [MHz] & [ms] & [ms] & [ms] & [ms] & [ms] & [ms] & [ms] & [ms] & [ms] & [ms] \\ 
\hline
20180924B & 362 & 0.3214 & 21.1 & 1297.5 & 0.864 & 2.0 & 0.59 & 8.64 & 3.85 & 56.5 & 0.55 & 7.01 & 0.41 & 5.31 \\
20181112A & 589 & 0.4755 & 19.3 & 1297.5 & 0.864 & 1.2 & 0.02 & 7.62 & 0.20 & 69.3 & 0.09 & 6.19 & 0.06 & 4.19 \\
20190102C & 365 & 0.29 & 14.0 & 1271.5 & 0.864 & 1.25 & 0.02 & 2.62 & 0.15 & 14.7 & 0.06 & 2.16 & 0.05 & 1.67 \\
20190608B & 340 & 0.1178 & 16.1 & 1269.5 & 1.728 & 10.8 & 3.83 & 19.8 & 13.8 & 72.1 & 1.57 & 15.5 & 1.40 & 13.9 \\
20190611B & 322 & 0.378 & 9.5 & 1252.7 & 1.728 & 1.59 & 0.03 & 0.72 & 0.19 & 4.70 & 0.06 & 0.56 & 0.04 & 0.41 \\
20190711A & 592 & 0.522 & 23.8 & 1172.9 & 1.728 & 10.9 & 0.00 & 24.2 & 0.05 & 161. & 8.59 & 21.5 & 5.65 & 14.1 \\
20190714A & 505 & 0.2365 & 10.7 & 1286.6 & 1.728 & 2.99 & 0.42 & 2.21 & 2.18 & 11.4 & 0.68 & 1.77 & 0.55 & 1.43 \\
20191001A & 507 & 0.23 & 37.1 & 826.4 & 1.728 & 13.4 & 4.52 & 26.4 & 3.92 & 22.9 & 0.45 & 20.8 & 0.36 & 16.9 \\
20191228B & 297 & 0.2432 & 22.9 & 1273.0 & 1.728 & 13.5 & 5.85 & 26.1 & 29.5 & 132. & 3.07 & 20.3 & 2.47 & 16.3 \\
20200430A & 380 & 0.161 & 13.9 & 863.5 & 1.728 & 22.6 & 6.5 & 24.7 & 5.65 & 21.5 & 7.58 & 20.0 & 6.53 & 17.2 \\
20200906A & 578 & 0.3688 & 16.1 & 846.4 & 1.728 & 0.12 & 0.03 & 17.3 & 0.04 & 22.8 & 0.04 & 14.1 & 0.03 & 10.3 \\
20210117A & 729 & 0.214 & 17.7 & 1274.5 & 1.182 & 3.58 & 0.25 & 10.0 & 1.18 & 47.3 & 1.20 & 8.31 & 0.98 & 6.84 \\
20210320C & 385 & 0.28 & 15.3 & 828.4 & 1.728 & 0.88 & 0.19 & 11.2 & 0.19 & 11.1 & 0.29 & 9.17 & 0.23 & 7.16 \\
20211127I & 235 & 0.0469 & 37.9 & 1272.5 & 1.182 & 0.48 & 0.02 & 18.0 & 0.07 & 54.2 & 0.22 & 14.6 & 0.21 & 14.0 \\
20211203C & 636 & 0.3439 & 14.2 & 891.4 & 1.182 & 25.4 & 1.66 & 10.1 & 2.54 & 15.5 & 12.2 & 14.6 & 9.10 & 10.8 \\
20211212A & 200 & 0.0707 & 10.5 & 1490.8 & 1.182 & 5.62 & 1.8 & 3.66 & 10.9 & 22.2 & 0.18 & 1.90 & 0.16 & 1.77 \\
20220105A & 583 & 0.2785 & 9.8 & 1649.8 & 1.182 & 2.25 & 0.43 & 1.12 & 6.65 & 17.4 & 0.79 & 1.08 & 0.61 & 0.85 \\
20220501C & 450 & 0.381 & 14.8 & 864.5 & 1.182 & 6.9 & 0.35 & 16.4 & 0.51 & 24.2 & 6.08 & 14.7 & 4.40 & 10.6 \\
20220610A & 1458 & 1.015 & 23.9 & 1149.4 & 1.182 & 2.0 & 0.52 & 18.0 & 7.43 & 257. & 0.85 & 14.7 & 0.42 & 7.31 \\
20220725A & 290 & 0.1926 & 10.9 & 1149.4 & 1.182 & 8.01 & 2.29 & 5.80 & 6.77 & 17.1 & 1.97 & 4.32 & 1.65 & 3.62 \\
20220918A & 643 & 0.491 & 26.3 & 1133.5 & 1.182 & 13.8 & 7.66 & 16.2 & 41.9 & 88.9 & 6.52 & 11.3 & 4.37 & 7.63 \\
20221106A & 343 & 0.2044 & 19.7 & 1649.6 & 1.182 & 6.89 & 0.18 & 16.8 & 2.35 & 218. & 5.32 & 14.6 & 4.42 & 12.2 \\
20230526A & 316 & 0.157 & 22.1 & 1272.2 & 1.182 & 2.7 & 1.16 & 16.3 & 4.70 & 66.4 & 1.40 & 13.3 & 1.21 & 11.5 \\
20230708A & 412 & 0.105 & 30.5 & 920.5 & 1.182 & 23.5 & 0.24 & 17.9 & 0.23 & 17.4 & 1.10 & 14.7 & 0.99 & 13.3 \\
20230718A & 477 & 0.035 & 10.9 & 1272.2 & 1.182 & 0.69 & 0.11 & 1.81 & 0.33 & 5.27 & 0.53 & 1.52 & 0.51 & 1.47 \\
20230902A & 440 & 0.3619 & 11.8 & 812.4 & 1.182 & 0.67 & 0.12 & 6.81 & 0.13 & 7.49 & 0.17 & 5.56 & 0.12 & 4.08 \\
20240201A & 375 & 0.0427 & 13.9 & 915.5 & 1.182 & 3.90 & 0.78 & 10.4 & 0.62 & 8.28 & 2.89 & 8.91 & 2.77 & 8.55 \\
20240210A & 284 & 0.0236 & 11.6 & 863.5 & 1.182 & 1.53 & 0.1 & 3.59 & 0.05 & 2.14 & 0.27 & 2.97 & 0.27 & 2.90 \\
20240310A & 602 & 0.127 & 19.1 & 846.4 & 1.182 & 13.4 & 2.23 & 17.6 & 1.63 & 12.9 & 3.17 & 14.5 & 2.81 & 12.8 \\
    \end{tabular}
    \caption{FRB properties used in this work: observed properties of FRB name, dispersion measure, redshift, and signal-to-noise ratio taken from \citet{Shannon_ICS}; high-time-resolution properties of signal-to-noise maximising width, and fitted scattering time taken from \citet{CRAFT_HTR}; and 
    maximum detectable scattering value, scattering time scaled to 1\,GHz in host rest-frame, and maximum detectable scattering time in host rest-frame; and intrinsic width, maximum detectable intrinsic width, intrinsic width in host rest frame, and maximum detectable width in host frame,
    derived in this work.}
    \label{tab:data}
\end{table*}

\end{document}